\documentstyle[11pt,pasp,twoside,psfig]{article}
\def\edcomment#1{\iffalse\marginpar{\raggedright\sl#1\/}\else\relax\fi}
\reversemarginpar

\begin{document}
\title{X-raying Active Galaxies with the New Generation of X-ray Observatories: Ionized
Outflows and High-Redshift Studies}
\author{W.N. Brandt, S.C. Gallagher, \& S. Kaspi}
\affil{Department of Astronomy \& Astrophysics, 525 Davey Laboratory, The Pennsylvania
State University, University Park, Pennslyvania 16802}
% 
% \author{Ima Co-Author}
% \affil{The Name of My Institution, The Full Address of My Institution}

\begin{abstract}
We briefly review X-ray studies of ionized AGN outflows and high-redshift AGN. 
We discuss recent progress with {\it Chandra\/} and {\it XMM-Newton\/} as well
as prospects and requirements for future X-ray missions. 
\end{abstract}

\section{Introduction}

The main goals of X-ray studies of active galactic nuclei (AGN) are (1) to measure the key 
parameters of the accreting black hole system, such as the black hole mass, spin, and
accretion rate, (2) to determine the distribution, dynamics and physical conditions of
nuclear and circumnuclear material (e.g., the accretion disk and its corona, the torus, 
jets, winds, clouds, and any circumnuclear starburst), and (3) to understand the cosmic
X-ray evolution of AGN. The talks and posters at this symposium have made it clear
that studies with {\it Chandra\/} and {\it XMM-Newton\/} have already touched all aspects
of AGN research, and it is wonderful to see such progress. 

This proceedings paper will briefly cover two important areas of AGN research that should
show exciting growth during the new century: (1) X-ray studies of ionized outflows in 
Seyfert galaxies and quasars, and (2) X-ray studies of the highest-redshift AGN. Only 
limited citations will be possible due to space limitations; our apologies in advance. 

\section{Ionized Outflows in Seyfert Galaxies and Quasars}

One of the most revolutionary aspects of the new observatories are the gratings spectra
they regularly provide; in AGN outflow research these have given qualitatively new
information by increasing the number of spectral features available for study from 
2--3 to 30--60 (see Figures~1 and 2). These features include absorption lines, emission 
lines, unresolved transition arrays, and edges. They provide constraints on the dynamics, 
geometry, and physical conditions in an AGN outflow. For example, the Doppler shifts 
of absorption lines have shown that the X-ray absorber is indeed outflowing with
typical bulk velocities of a few hundred km~s$^{-1}$ and can have a velocity dispersion
comparable to the bulk velocity (e.g., Kaspi et~al. 2001 and references therein). 
In some cases multiple velocity components and P~Cygni profiles are discernible. 
In terms of geometry, X-ray emission-line strengths confirm that these outflows
have large covering factors ($\approx$~0.3--0.8), but the characteristic distance
from the black hole and radial extent are still poorly constrained. Pinning down
these quantities will require a better understanding of the physical conditions
in the absorbing gas, particularly its density. It is also crucial to determine
if the gas is in photoionization equilibrium and to assess its abundances and 
dust content. Hopefully the long gratings observations currently being performed
will provide the requisite constraints. 

\begin{figure}[t!]
\hspace{0.8cm} \psfig{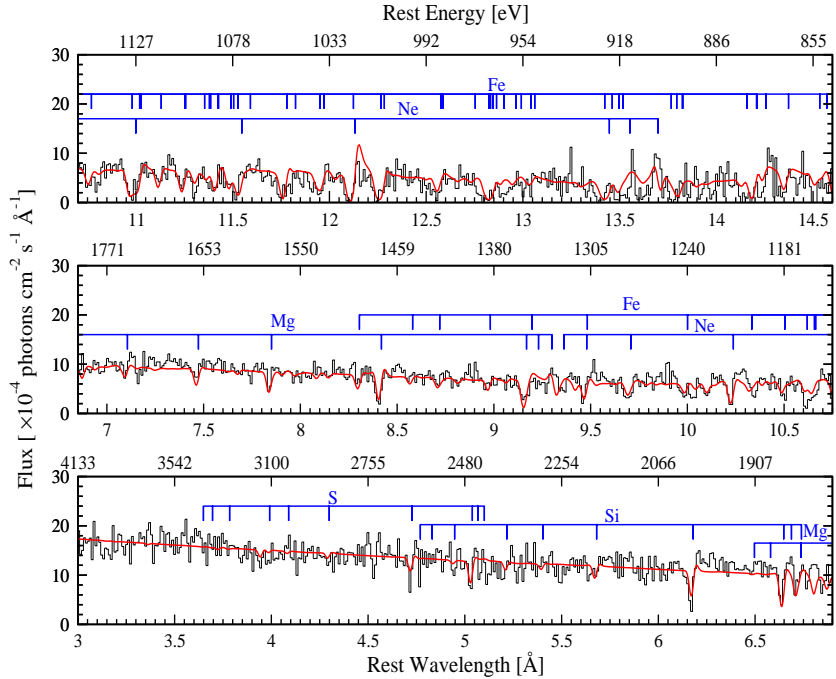}
\caption{Part of the Cycle~1 {\it Chandra\/} HETGS spectrum of the bright Seyfert~1
galaxy NGC~3783. The histogram shows the data, and the smooth curve shows a 
photoionization model for the data. The H-like and He-like lines of 
Ne, Mg, Si, and S as well as the strongest lines contributing to the model
are marked. From Kaspi et~al. (2001).}
\end{figure}

\begin{figure}[t!]
\hspace{1.2cm} \psfig{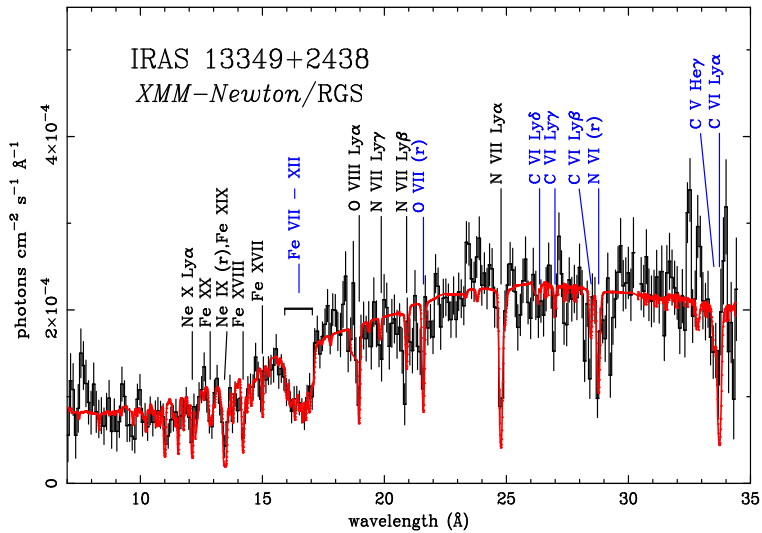}
\caption{The AO1 {\it XMM-Newton\/} RGS spectrum of the infrared-loud
quasar IRAS~13349+2438. The histogram shows the data, and the smooth curve
shows a model for the data. The strongest observed lines are marked; 
note the unresolved transition array from $\approx$~16--17~\AA\ that explains
some of the previously detected spectral complexity. From Sako et~al. (2001).}
\end{figure}

Aside from the measurement of key parameters, X-ray gratings studies have also
allowed a re-evaluation of the basic interpretation of the low-energy spectral
complexity seen from many Seyfert galaxies (e.g., Branduardi-Raymont et~al. 2001; 
Lee et~al. 2001); some have argued for the presence of accretion-disk lines 
formed close to a Kerr black hole. The current debate over these lines should 
be resolvable with better spectral data and constraints on the rapid variability 
of the low-energy spectral complexity.

Looking further into the future of X-ray outflow studies, it is important that
X-ray grating spectrometers achieve better spectral resolution (also see the
paper by M.~Elvis in these proceedings). The current ones, while a {\it major\/}
advance, have velocity resolutions comparable to that of {\it IUE\/} 
($\approx$~300--600~km~s$^{-1}$). Just as {\it HST\/} revealed unresolved 
structure (e.g., multiple velocity components) in many of the ultraviolet 
lines studied by {\it IUE\/}, future X-ray missions may further subdivide 
the currently detected X-ray absorption lines (see Figure~3). 
Larger collecting area is also of great importance; currently even 
the brightest Seyfert galaxies require gratings observations of $\ga 1$~day, and X-ray 
gratings studies of the more powerful outflows known in luminous quasars
[e.g., Broad Absorption Line (BAL) quasars] are essentially impossible. 
The few X-ray CCD spectra available for BAL quasars suggest that X-ray BALs
may be awaiting discovery (e.g., Gallagher et~al. 2001; Mathur et~al. 2001). 
Given the large X-ray column densities observed in BAL quasars, constraints 
on the velocity structure of the X-ray absorbing gas will have fundamental 
implications for our understanding of the energy budget of luminous AGN.

\begin{figure}[t!]
\hspace{0.2cm} \psfig{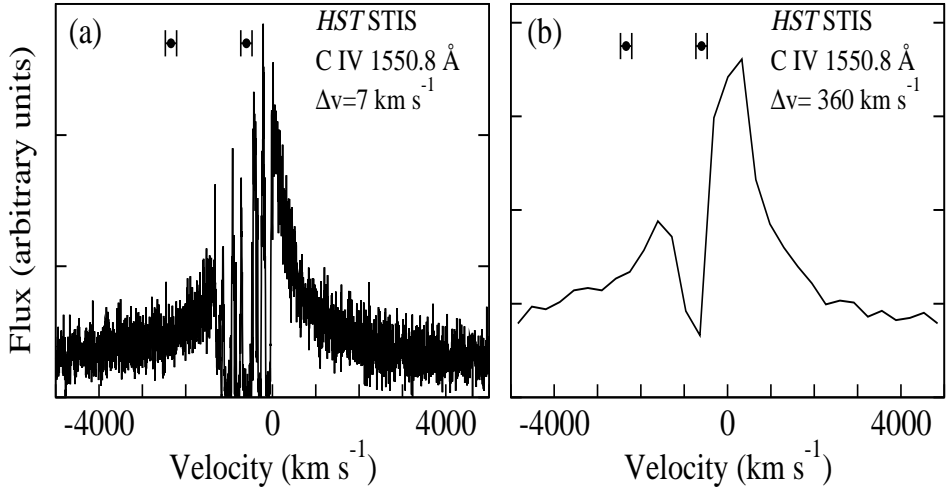}
\caption{{\it HST\/} STIS velocity spectra of the C~{\sc iv} 1548.2~\AA\ and
1550.8~\AA\ emission and absorption lines seen from the bright Narrow-Line 
Seyfert~1 galaxy NGC~4051. Panel (a) shows the lines at full resolution;
the ``scruffy'' structure from $\approx -1000$~km~s$^{-1}$ to $\approx 0$~km~s$^{-1}$ 
is due to the presence of at least nine kinematically distinct absorption 
components. When the full-resolution spectrum is smoothed to the approximate velocity 
resolution of the {\it Chandra\/} HETGS in panel (b), the kinematic components 
are not visible and the resulting absorption ``line'' resembles the lines seen 
in X-rays by the {\it Chandra\/} HETGS. The X-ray absorbers seen in 
Seyfert galaxies may be subdivided into further systems that cannot be resolved
with current X-ray instruments. 
The two dots with error bars near the tops of the panels show the 
velocities of the X-ray absorption components that {\it can\/} be resolved
by {\it Chandra\/}; note that only one of the two X-ray absorption 
components shows corresponding ultraviolet absorption. 
From Collinge et~al. (2001).}
\end{figure}

\section{X-rays from the Dawn of the Modern Universe}

One of the main themes in astronomy over the coming decades will be the 
exploration of the dawn of the modern Universe, when the first stars, 
galaxies, and black holes formed, ending the cosmic ``dark age''
(e.g., the USA National Research Council 2000 Decadal 
Report). X-ray astronomy can play a crucial role in this project by allowing
studies of warm and hot objects in the early Universe, thereby complementing
studies of the cooler Universe with observatories such as {\it NGST\/}, {\it ALMA\/}, 
{\it Herschel\/}, {\it SIRTF\/}, and {\it SKA\/}. Large X-ray missions such
as {\it XEUS\/} ($\approx \$2$ billion) are being planned to focus on the
first massive black holes (at $z\approx$~5--20) and black hole evolution. 

\begin{figure}[t!]
\hspace{2.2cm} \psfig{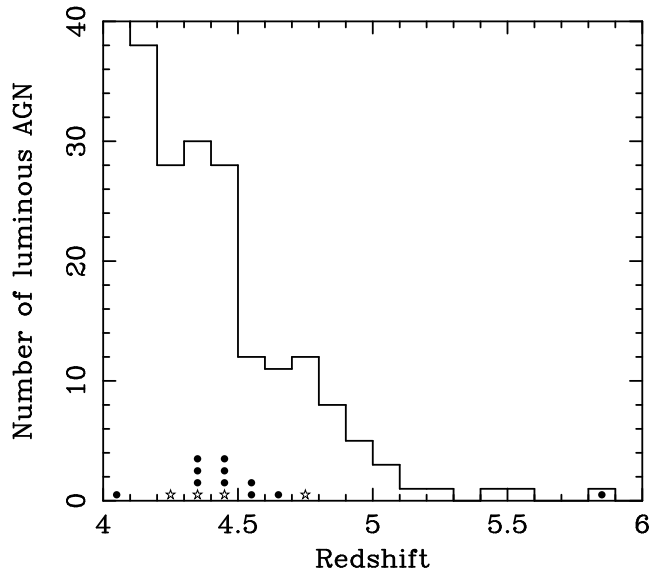}
\caption{Histogram of the redshift distribution of published luminous AGN
at $z>4$. The stars denote blazars detected in X-rays, and the solid dots
denote quasars and a Seyfert galaxy detected in X-rays.}
\end{figure}

It is critical to use {\it Chandra\/} and {\it XMM-Newton\/} to lay the 
observational groundwork for future high-redshift X-ray efforts. At present,
our knowledge about the $z>4$ X-ray Universe is unfortunately quite limited. 
Only 16 objects at $z>4$ have published X-ray detections: 10 quasars, 
four blazars, one Seyfert galaxy, and one gamma-ray burst afterglow. The 
quasars and Seyfert galaxy have limited counts (typically 10--50); their 
broad-band spectral energy distributions look roughly consistent with those 
of lower-redshift objects (e.g., Kaspi, Brandt, \& Schneider 2000; 
Brandt et~al. 2001a). The four blazars are the only $z>4$ objects with 
X-ray spectra. Three of them show low-energy cutoffs apparently due to 
X-ray absorption (e.g., Fabian et~al. 2001 and references therein), 
extending the absorption-redshift trend found for radio-loud quasars at 
lower redshifts (e.g., Fiore et~al. 1998). 

Figure~4 shows the redshift distribution of published luminous AGN 
at $z>4$ with published X-ray detections marked. Most of the X-ray 
detections are at $z<4.6$, and there is 
only one at $z>5$ separated from the others by $\Delta z>1$. 
This object, SDSS~1044--0125 at $z=5.80$, was recently detected by 
{\it XMM-Newton\/} and was found to be $\approx 10$ times weaker in
X-rays than expected given its optical flux (Brandt et~al. 2001b). 
Absorption was proposed as the most likely cause of its X-ray weakness,
and BALs have indeed recently been discovered by Maiolino et~al. (2001)
and F.H. Chaffee et~al., in preparation. Given that SDSS~1044--0125 
is likely to be an unusual member of the $z>5$ population, at present we 
have {\it no\/} constraints on the X-ray properties of normal $z>5$ quasars. 
X-ray astronomers are largely ``groping in the dark (age)'' when attempting to 
plan missions to study AGN at $z\approx$~5--20. Fortunately, over the 
next five years surveys such as the Sloan Digital Sky Survey (SDSS; 
York et~al. 2000) should provide a large number of $z=$~5--6.5 AGN, 
many of which should be bright enough for study in X-rays. It should 
also be possible to detect moderate-luminosity AGN at $z=$~5--10
in the deepest X-ray surveys (e.g., Haiman \& Loeb 1999); follow-up
of such objects will prove challenging. Moderate-to-extreme photon 
starvation will limit our ability to perform X-ray spectroscopy of the 
highest-redshift AGN for the next few years, even with {\it XMM-Newton\/}
(barring fortuitous finds such as lensed quasars or bright blazars). 
However, prospects are good for {\it Constellation-X\/} and 
{\it XEUS\/} which should be able to deliver many 10,000--100,000
count spectra of the highest redshift AGN (see Figure~5). 

\begin{figure}[t!]
\hspace{1.6cm} \psfig{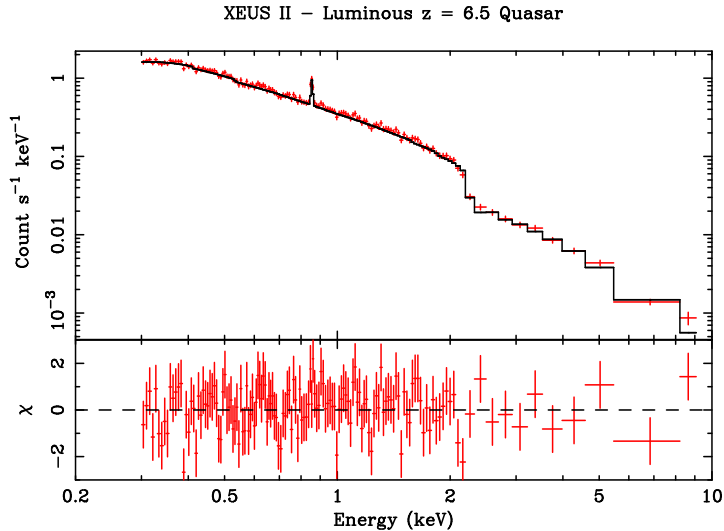}
\caption{Simulated 40~ks {\it XEUS\/} (final configuration) spectrum of a luminous
$z=6.5$ quasar such as might be found by the SDSS. The quasar has a power-law 
spectrum with a photon index of $\Gamma=2$, a Galactic column density of 
$N_{\rm H}=2\times 10^{20}$~cm$^{-2}$, and an observed-frame 0.5--2.0~keV flux 
of $4\times 10^{-15}$~erg~cm$^{-2}$~s$^{-1}$. A narrow iron~K$\alpha$ line at 
6.4~keV is included with a rest-frame equivalent width of 150~eV.}
\end{figure}

\acknowledgments

We thank the organizers for an excellent symposium. We gratefully acknowledge 
the financial support of 
NASA grant NAS8-38252 (WNB, SK, SCG),
NASA LTSA grant NAG5-8107 (WNB, SK), 
CXC grant GO0-1160X (WNB, SK), and
NASA GSRP grant NGT5-50277 (SCG).

\end{document}